\documentclass[aps,prc,reprint,superscriptaddress,showpacs,nofootinbib]{revtex4-1}
\usepackage{graphicx}
\usepackage{subfigure}
\usepackage{default}
\usepackage{longtable}
\usepackage{xfrac}
\newcommand{\textunderscript}[1]{$_{\text{#1}}$}

\begin{document}
\title{Systematic investigation of projectile fragmentation using beams of unstable B and C isotopes}
%

%
\author{R.~Thies}
\email{ronja.thies@chalmers.se}
\affiliation{Institutionen f\"or Fysik, Chalmers Tekniska H\"ogskola, 412 96 G\"oteborg, Sweden}

\author{A.~Heinz}
\affiliation{Institutionen f\"or Fysik, Chalmers Tekniska H\"ogskola, 412 96 G\"oteborg, Sweden}

\author{T.~Adachi}
\affiliation{KVI, University of Groningen, Zernikelaan 25, 9747 AA Groningen, The Netherlands}

\author{Y.~Aksyutina}
\affiliation{Institut f\"ur Kernphysik, Technische Universit\"at Darmstadt, 64289 Darmstadt, Germany}
\affiliation{GSI Helmholtzzentrum f\"ur Schwerionenforschung, 64291 Darmstadt, Germany}

\author{J.~Alcantara-N\'u\~nes}
\affiliation{Dpt. de F\'{i}sica de Part\'{i}culas, Universidade de Santiago de Compostela, 15706 Santiago de Compostela, Spain}

\author{S.~Altstadt}
\affiliation{Goethe-Universit\"at Frankfurt am Main, 60438 Frankfurt am Main, Germany}

\author{H.~Alvarez-Pol}
\affiliation{Dpt. de F\'{i}sica de Part\'{i}culas, Universidade de Santiago de Compostela, 15706 Santiago de Compostela, Spain}

\author{N.~Ashwood}
\affiliation{School of Physics and Astronomy, University of Birmingham, Birmingham B15 2TT, United Kingdom}

\author{T.~Aumann}
\affiliation{Institut f\"ur Kernphysik, Technische Universit\"at Darmstadt, 64289 Darmstadt, Germany}
\affiliation{GSI Helmholtzzentrum f\"ur Schwerionenforschung, 64291 Darmstadt, Germany}

\author{V.~Avdeichikov}
\affiliation{Department of Physics, Lund University, 22100 Lund, Sweden}

% ...

\collaboration{R3B collaboration}
\noaffiliation

%
%


\begin{abstract}
\begin{description}
\item[Background] Models describing nuclear fragmentation and fragmentation-fission deliver important input for planning nuclear physics experiments and future radioactive ion beam facilities. These models are usually benchmarked against data from stable beam experiments. In the future, two-step fragmentation reactions with exotic nuclei as stepping stones are a promising tool to reach the most neutron-rich nuclei, creating  a need for models to describe also these reactions. %
\item[Purpose] We want to extend the presently available data on fragmentation reactions towards the light exotic region on the nuclear chart. Furthermore, we want to improve the understanding of projectile fragmentation especially for unstable isotopes.%
\item[Method] We have measured projectile fragments from \textsuperscript{10,12-18}C and \textsuperscript{10-15}B isotopes colliding with a carbon target. These measurements were all performed within one experiment, which gives rise to a very consistent dataset. We compare our data to model calculations.%
\item[Results] One-proton removal cross sections with different final neutron numbers (1pxn) for relativistic \textsuperscript{10,12-18}C and \textsuperscript{10-15}B isotopes impinging on a carbon target. Comparing model calculations to the data, we find that EPAX is not able to describe the data satisfactorily. Using ABRABLA07 on the other hand, we find that the average excitation energy per abraded nucleon needs to be decreased from 27 MeV to 8.1 MeV. With that decrease ABRABLA07 describes the data surprisingly well.
\item[Conclusions] Extending the available data towards light unstable nuclei with a consistent set of new data have allowed for a systematic investigation of the role of the excitation energy induced in projectile fragmentation. Most striking is the apparent mass dependence of the average excitation energy per abraded nucleon. Nevertheless, this parameter, which has been related to final-state interactions, requires further study.

\end{description}
\end{abstract}

\maketitle

\section{Introduction}
Since the advent of radioactive ion beam facilities it is possible to study more exotic isotopes, which has led to new discoveries, like halo-nuclei \cite{ref:riisager2013symposium} and the changing of magic numbers with isospin \cite{ref:Kanungo2013symposium}. Reaction cross sections involving exotic nuclei allow us to extract nearly model-independent observables, in contrast to other reaction processes such as nucleon transfer which is strongly dependent on the reaction mechanism adopted for the experimental analysis. Indeed, reaction cross sections have led to a number of interesting discoveries such as the above mentioned halo-nuclei \cite{ref:bonaccorso2013symposium}.\\
Models describing nuclear fragmentation and fragmentation-fission deliver important input to yield predictions, useful for planning of experiments and future accelerator facilities \cite{ref:summerer2012}. Recently, two-step fragmentation reactions have been discussed for future facilities \cite{ref:nilsson2013} and are already used \cite{ref:perez2011doublefrag} to reach especially neutron-rich nuclei.\\
There exist several models for the prediction of reaction cross sections, examples are models following the  abrasion-ablation, the intra-nuclear cascade approach and empirical parametrizations. 
As the models are usually benchmarked with stable nuclei -- while exotic nuclei can exhibit different behaviour -- their ability to predict fragmentation cross sections for exotic nuclei is unclear. 
 We investigate whether fragmentation models are able to describe reaction cross sections of light exotic nuclei, which exhibit such a rich variety of properties.\\
 We have systematically measured one-proton-x-neutron (1pxn) removal cross sections for 0 $\leq$ x $\leq$ 5 for a large range of carbon and boron isotopes impinging on carbon targets at relativistic energies.
We compare our measured 1pxn removal cross sections to calculations of an abrasion-ablation model (ABRABLA07 \cite{ref:gaimard1991}). We also compare to the widely used EPAX code \cite{ref:suemmerer2013epax} though it is limited to $A > 40$, since it has been used earlier for lighter nuclei. Leistenschneider \textit{et al.} \cite{ref:leistenschneider2002} performed a similar study for the less exotic \textsuperscript{17-21}O isotopes, comparing both models to their data. The comparison was unsatisfactory, but subsequently both models have been improved.
\section{Experiment}

The experiment was conducted using the LAND/R\textsuperscript{3}B set-up at the GSI Helmholtz Centre for Heavy Ion Research in Germany, and was designed as an overview experiment covering isotopes with Z=3 to Z=9 between the extremes of isospin.\\
\begin{figure}
\includegraphics[width=\columnwidth]{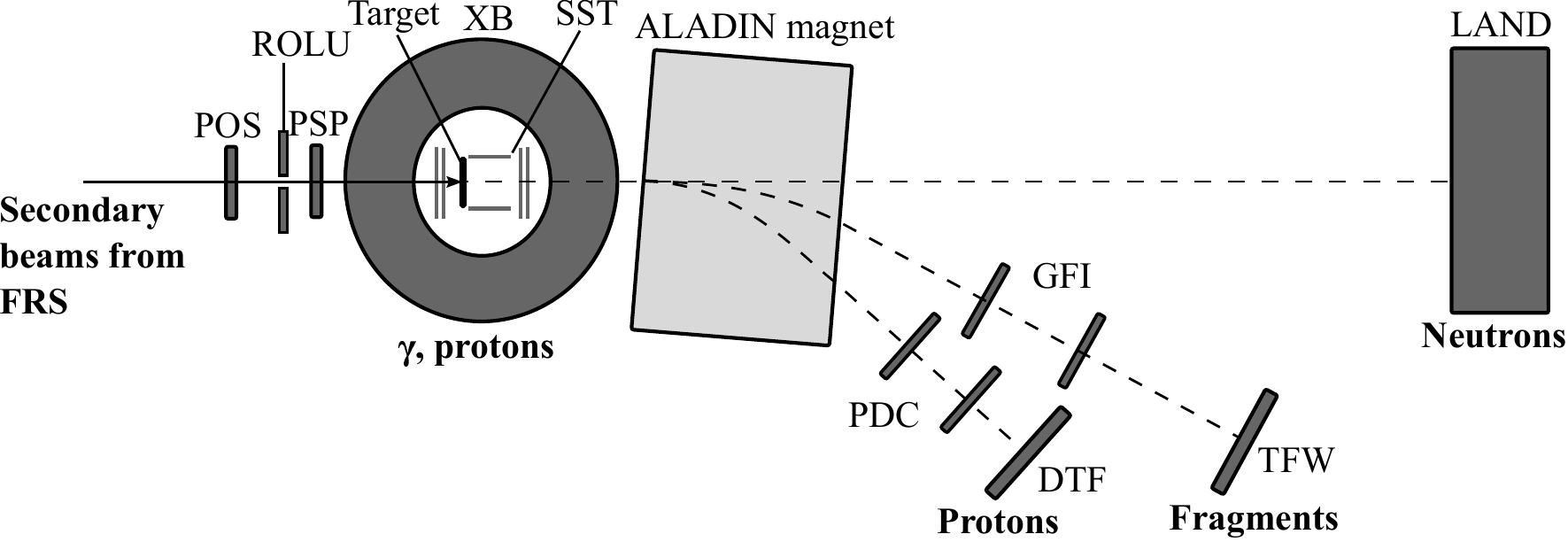}
\caption{Schematic view of the LAND/R\textsuperscript{3}B set-up seen from above.  The most important detectors for this work are POS, ROLU, PSP, SST, GFI, TFW, and XB. POS provides energy-loss ($\Delta E$) and time-of-flight (ToF) measurements. ROLU is an active veto detector on the incoming beam. PSP and SST are used for $\Delta E$ measurements, the main purpose of the SST is to determine incoming and outgoing directions of the beam. The GFI provide tracking of the beam behind the magnet ALADIN, and the TFW provides ToF, $\Delta E$ and position information. The XB is a calorimeter for protons and $\gamma$'s, and are here solely used for trigger purposes. For a more detailed description of the set-up see text. This schematic is not to scale.}
\label{fig:setup}
\end{figure}
The radioactive beams were produced from an \textsuperscript{40}Ar primary beam at 490 $A$MeV\footnote{Here, and in all further uses of the unit $A$MeV, we neglect the binding energy.} impinging on a 4 g/cm\textsuperscript{2} Be target. To separate and select the secondary beams the FRS fragment separator \cite{ref:geissel1992FRS} was used. With 5 different separator settings, beams with (centred) $A/Z$ ratios ranging from 1.66 to 3 were selected and guided to the experimental set-up. The secondary beams had kinetic energies in the range from 390 $A$MeV to 430 $A$MeV. 
Reaction targets of C (0.56 $\mathrm{g~cm^{-2}}$ and 0.93 $\mathrm{g~cm^{-2}}$) as well as an empty target frame were used in this work.\\
The LAND/R\textsuperscript{3}B set-up, shown in Fig.~\ref{fig:setup}, is designed for complete kinematics measurements on an event-by-event basis. At relativistic beam energies, the set-up benefits from kinematic forward-focusing of the reaction products, resulting in almost full acceptance in the centre-of-mass frame. The incoming ions are characterized by their magnetic rigidity (defined by the FRS), by their ToF (Time-of-Flight) between the FRS and the set-up measured by plastic scintillator detectors (POS), and by energy-loss measurements ($\Delta E$) in a silicon PIN diode (PSP) upstream from the reaction target. Located directly in front of and behind the reaction target are pairs of double-sided-silicon-strip detectors, SST1 through SST4, (100 $\mu$m pitch) determining the angle and charge of incoming and outgoing ions.\\
Light reaction products emitted at lab-angles $>$ 7.5$^{\circ}$ are detected in the segmented NaI array Crystal Ball (XB) \cite{metag1983XB} surrounding the target. By means of a dual read-out in the forward direction \cite{felix_phd} (up to 63$^{\circ}$ from the beam direction) the array is capable of detecting both photons and protons emitted at large angles, though with limited angular precision ($\approx$ 77 msr solid angle per segment).\\
 Charged fragments are bent by the dipole magnet ALADIN and subsequently detected in fibre detectors (GFI) \cite{ref:Mahata2009GFI} for position determination in the bending plane. After a total flight-path of around 10 m behind the target, the fragments are detected in TFW, a plastic ToF wall providing time, energy-loss and coarse position information. \\
Beam-like protons emitted at small angles ($<$ 7.5$^{\circ}$) also traverse the magnet and are detected by two drift chambers (PDC) and a ToF wall (DTF).
 Neutrons (emitted at angles $<$ 7.5$^{\circ}$) are detected in forward direction, about 12 m downstream from the target in the neutron detector LAND  \cite{ref:blaich1992LAND}.
The data presented in this work do not require reconstruction of neutrons and light reaction products. Though the set-up also allows detailed spectroscopic analysis, this is not within the scope of this work. Cross section measurements require significantly less statistics, and therefore allow for an overview of all ions in the experiment (we restrict ourselves here to boron and carbon).

\section{Analysis}
\begin{figure}
    \includegraphics[width=\columnwidth]{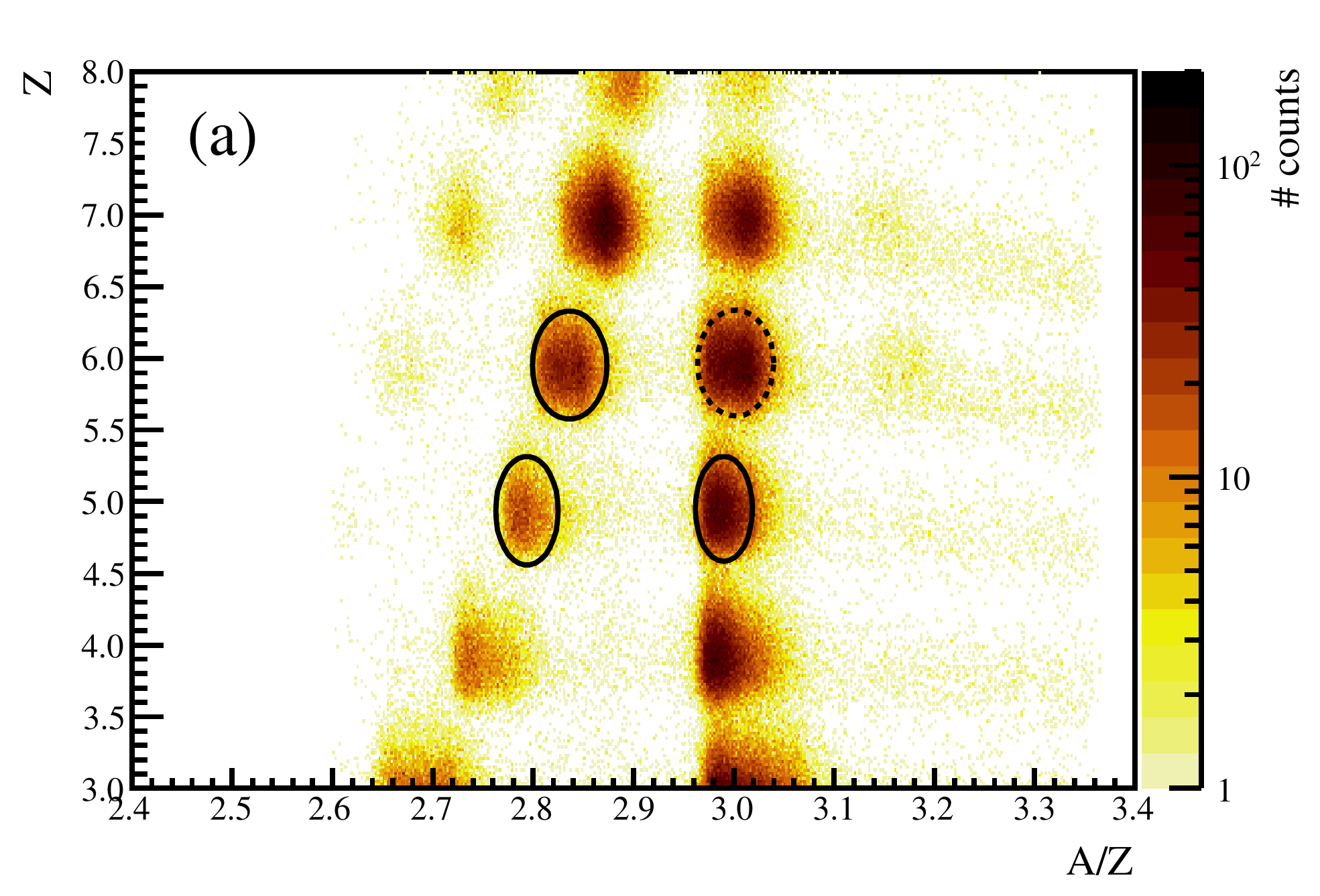}
    \newline
  \includegraphics[width=\columnwidth]{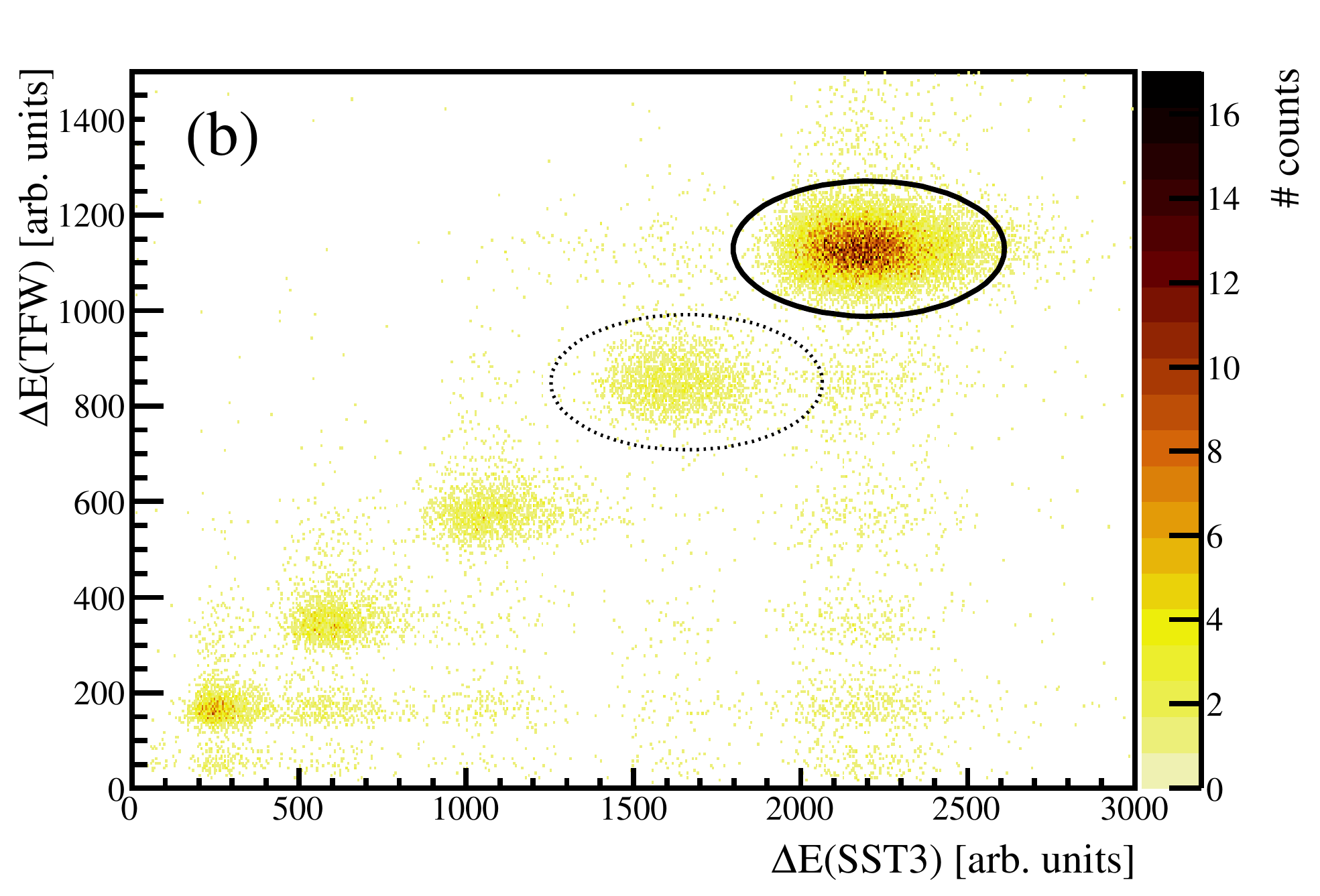}
  \newline
    \includegraphics[width=\columnwidth]{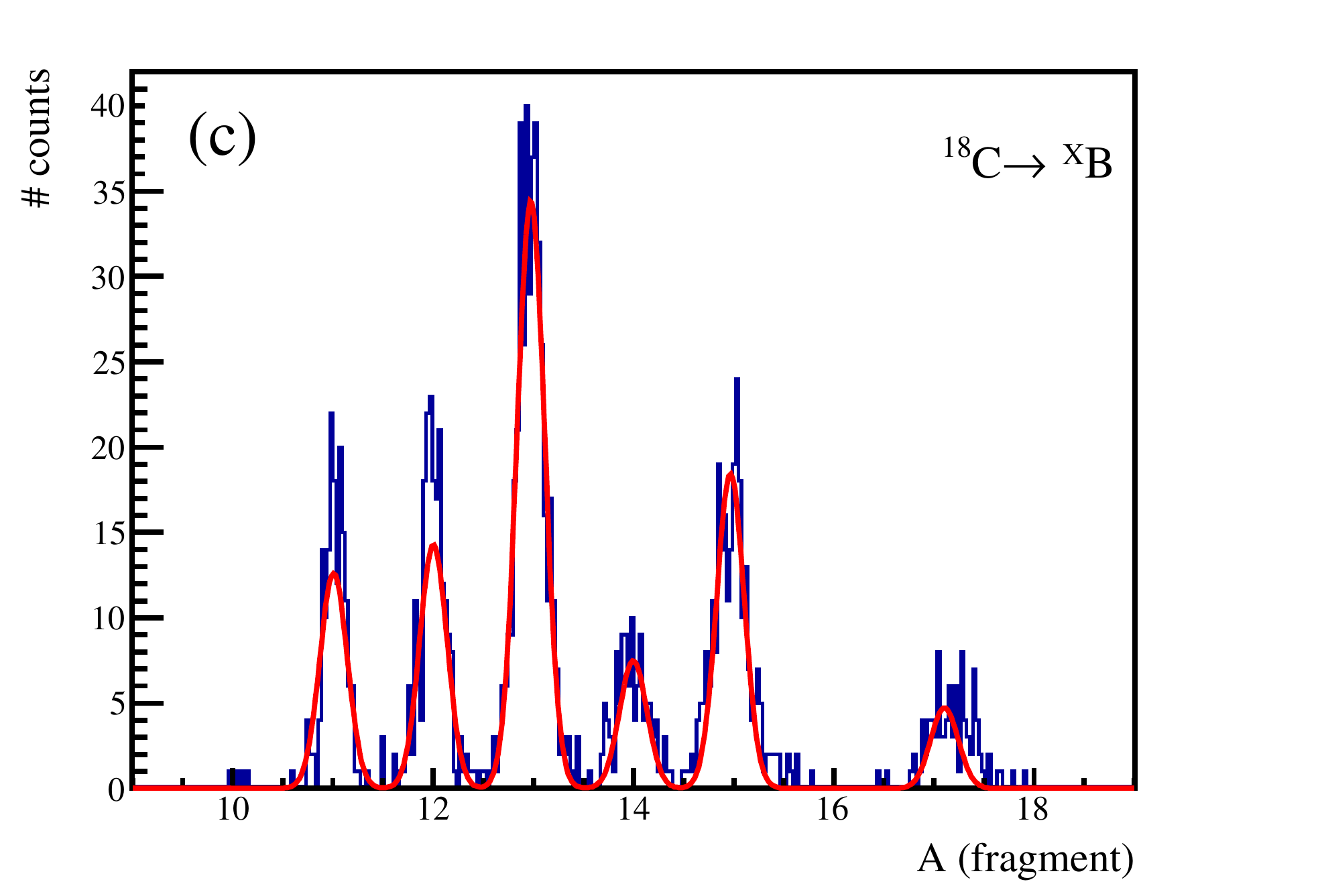}
\caption{(Colour online) Illustration of the reaction identification (ID). (a) shows the incoming ID with charge versus mass-to-charge ratio. The ellipses indicate the 2$\sigma$ selection of different isotopes. The dashed ellipse represents the selection used for the data in the plots (b) and (c). (b) presents the charge identification after the reaction target, using $\Delta$E measurements at the end of the set-up versus in the first detector behind the reaction target. The ellipses indicate the 3$\sigma$ selection of the unreacted beam (solid) and 1pxn reaction (dotted). (c) shows the reconstructed mass from the 1pxn removal and the fit to the spectrum. For details see text.}
\label{fig:identification}
\end{figure}

The incoming beam is selected\footnote{In order to assure reproducibility: for calibration and unpacking the \textit{land02} software package with the following git-tags was used: ronja-r3bm-5-2015 (\textit{land02}) and ronja-6-2015 (calibration parameters). } by fitting the charge versus mass-to-charge-ratio distribution (see Fig.~\ref{fig:identification}(a)) with 2-dimensional (2D) Gaussian distributions.  Only ions inside the 2$\sigma$ selection around the mean value, extracted from the fit, are taken into account in the analysis. To further reduce misidentifications arising from pile-up, a second additional charge identification using $\Delta$E measurements from POS and the SST detector just upstream from the target is employed, following the same pattern; fitting of 2D Gaussian distributions and selecting ions inside 2$\sigma$ from the mean.\\
The charge of the outgoing ion is identified by using $\Delta$E measurements in the SST detector directly downstream from the target (SST3) and the ToF detector at the end of the set-up (TFW), thus ensuring that no charge-changing reactions take place while the fragment travels through the set-up behind the target, see Fig.~\ref{fig:identification}(b). The same technique of 2D-Gaussian distribution fits, but now with a 3$\sigma$ selection is used.\\
The mass of the outgoing fragment is calculated using the map of the magnetic field of ALADIN, the direction of the ion after the target, the direction after the magnet and the time of flight through the set-up using $\chi^2$ minimization of a Runge-Kutta propagation\footnote{In order to assure reproducibility: \textit{LAND/R\textsuperscript{3}B tracker} software was used with the git-tag ronja-r3bm-5-2015.} \cite{ref:tracker} of the ion through the set-up. An example of the resulting mass distribution for a 1pxn removal reaction is presented in Fig.~\ref{fig:identification}(c). We employ a fit of a sum of Gaussian distributions (where the number of distributions in the sum corresponds to the number of different isotopes produced) to these mass distributions, and extract the number of outgoing ions of a certain isotope using the fit-parameters.
Isotopes with cross sections below $\approx$2 mb do not have sufficient statistics, thus no cross sections are reported. Due to acceptance limits, no cross sections for neutron-loss channels with more than five neutrons ($\Delta N>5$) could be extracted.\\
The cross sections are normalized using the unreacted beam, which is identified and reconstructed in the same way as the reacted beam. Together with the $\Delta N\leq 5$ condition, ensuring that the fragment is inside the acceptance of our set-up, this renders efficiency corrections for beam-detectors unnecessary.\\
Two different trigger patterns\footnote{A trigger pattern is a certain combination of detectors firing, it is used for selecting which events are recorded.} are used in this analysis. For selection of the unreacted beam, the ``fragment-trigger" which requires valid ToF signals and no veto of the incoming beam (c.f. Fig.~\ref{fig:setup} ROLU). For the reacted beam a ``XB-reaction-trigger" was used, requiring in addition to the same conditions as the fragment-trigger, also the detection of an energy signal in the calorimeter surrounding the target (XB). The calorimeter detects $\gamma$-rays and light particles at angles $\geq 7.5^{\circ}$ with respect to the beam axis. An energy signal in the XB indicates therefore that a reaction took place. The trigger efficiency of the XB-reaction-trigger is experimentally determined to be (85.3 $\pm$ 2.5)\% of the trigger efficiency of the fragment-trigger.\\
The reaction probability of the carbon and boron isotopes in the carbon targets is (0.9 $\pm$ 0.2)\% and (0.8 $\pm$ 0.2)\% for the thinner and 
(1.5 $\pm$ 0.3)\% and (1.3 $\pm$ 0.3)\% for the thicker target, respectively. The probability of multiple reactions in the target is thus insignificant.
\section{Results}
\begin{figure}
\includegraphics[width=0.98\columnwidth]{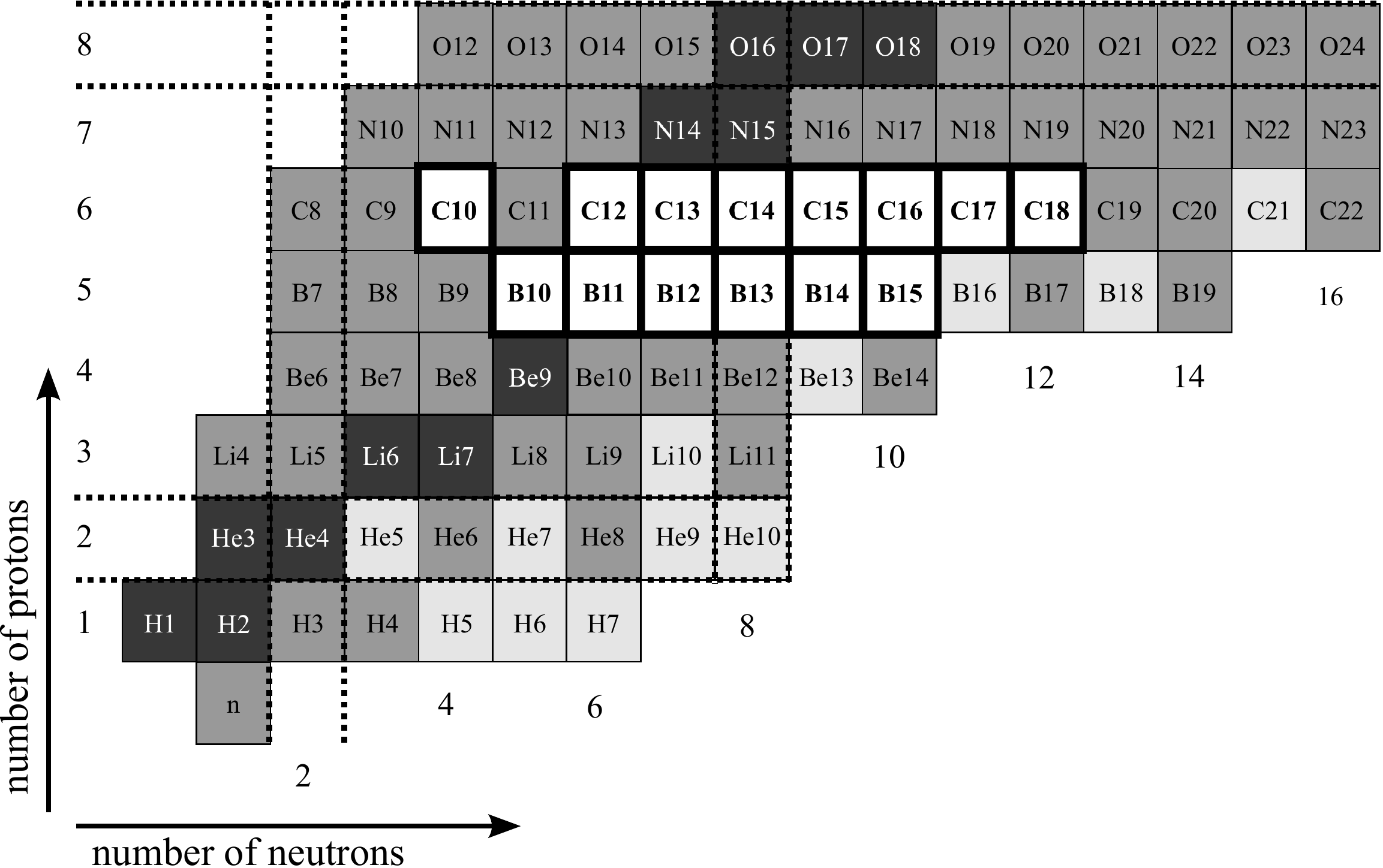}
\caption{Excerpt from the nuclear chart, illustrating the isotopes selected from the incoming secondary beams (white, thick frame). All carbon and boron isotopes with sufficient statistics were used. }
\label{fig:nuclearchart}
\end{figure}
\begin{figure}
\includegraphics[width=\columnwidth]{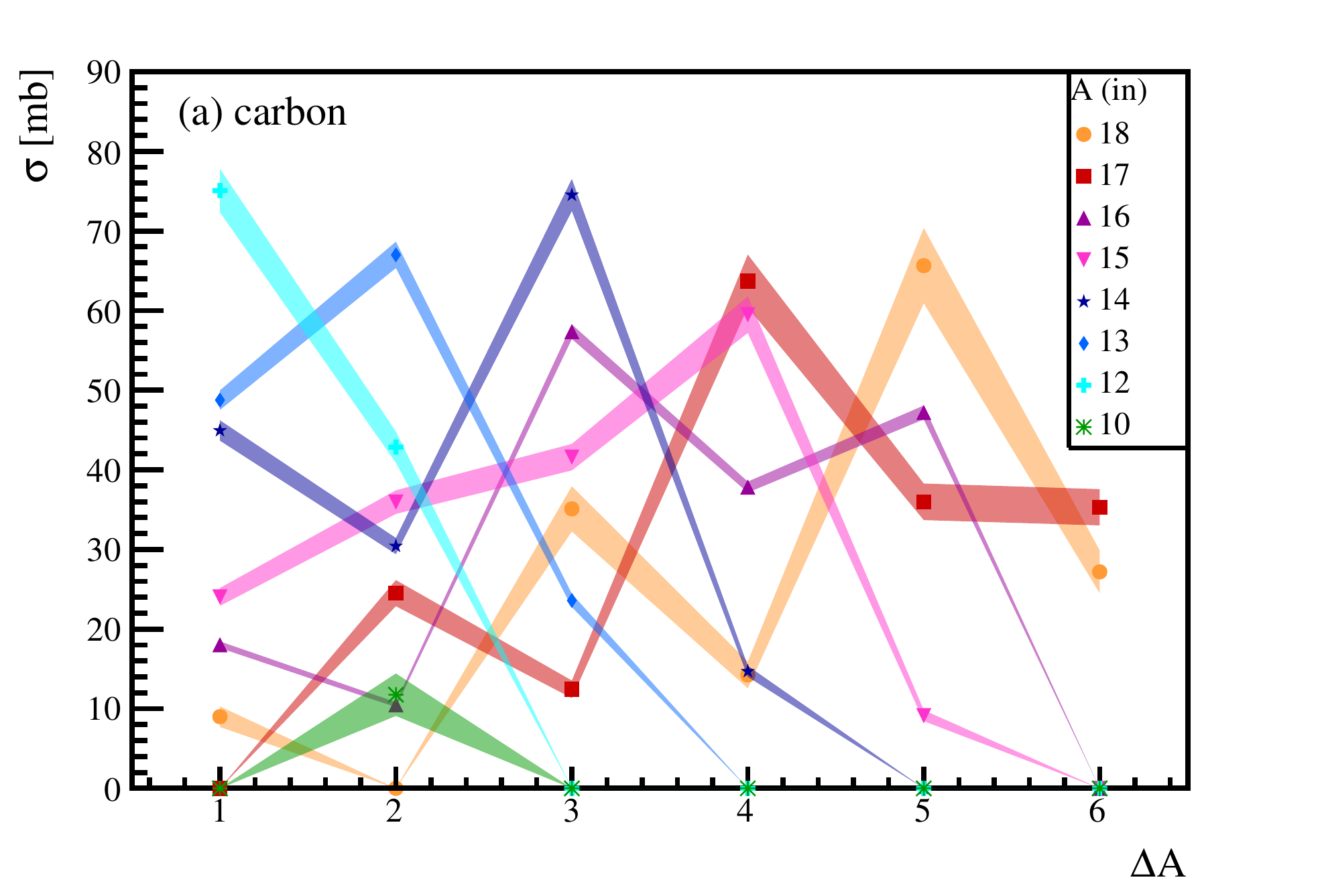}
\includegraphics[width=\columnwidth]{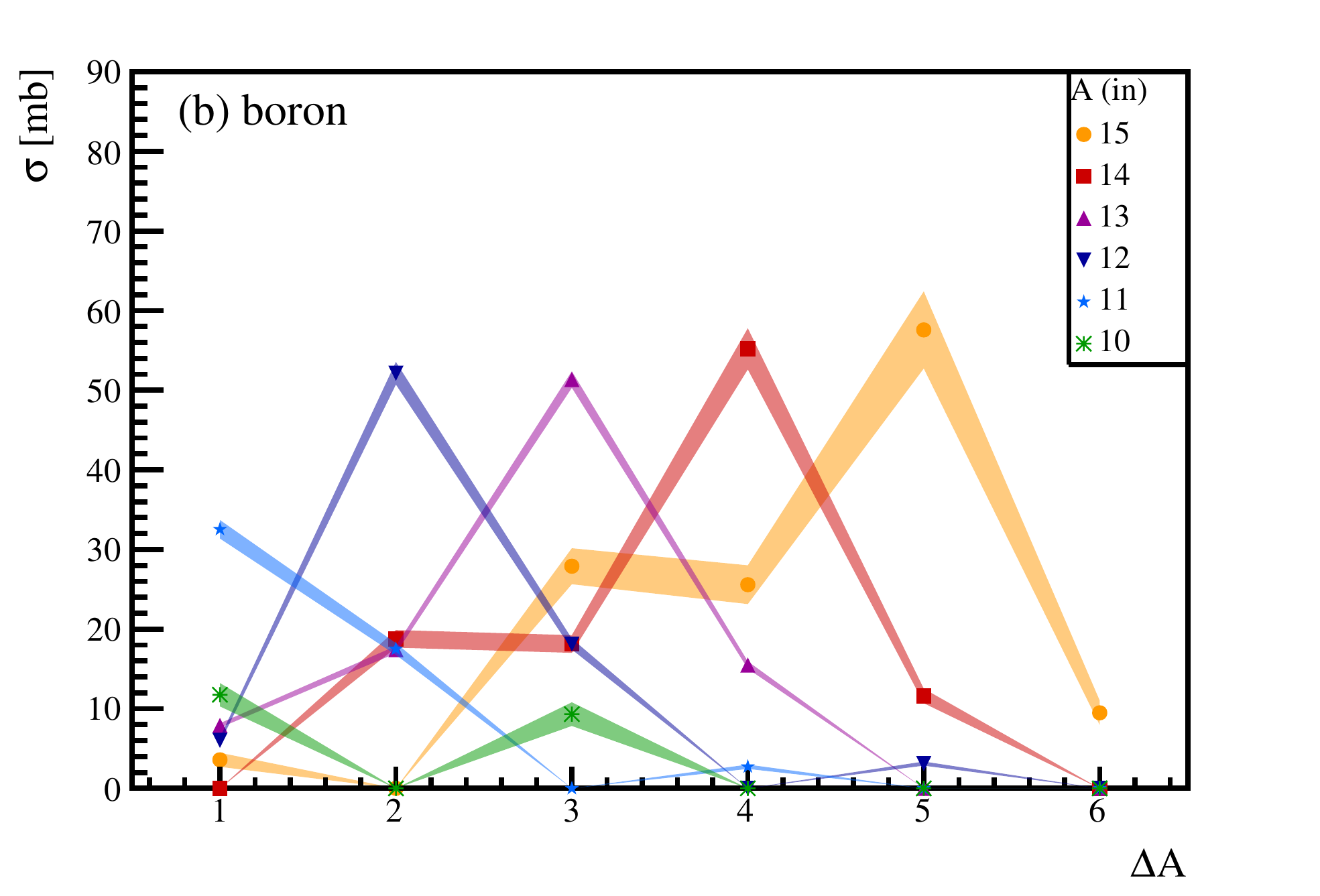}
\caption{(Colour online) 1pxn removal cross sections plotted versus the change in nucleon number, for carbon and boron. The shaded area represents the statistical error bar. For boron there is a strong trend that the cross section for populating the  long-lived \textsuperscript{10}Be is largest for all incoming isotopes. For carbon isotopes the cross section to produce the heaviest available stable isotope, \textsuperscript{11}B is largest, except for very neutron-rich isotopes, where instead the cross section to the semi-magic \textsuperscript{13}B becomes largest, with the transition point located at \textsuperscript{16}C.}
\label{fig:xsecoverview}
\end{figure}

We have extracted one-proton-x-neutron-removal (1pxn) cross sections for 0 $\leq$ x $\leq$ 5 for beams of carbon isotopes of mass 10 and 12 through 18, and boron isotopes of mass 10 through 15 on a C target. The location of these isotopes on the nuclear chart is illustrated in Fig.~\ref{fig:nuclearchart}.\\
\begin{table}
\begin{tabular}{c c c c c @{\hskip 6pt}|@{\hskip 6pt} c c c c c}
 \hline
 \hline
A\textunderscript{in}	&Z\textunderscript{in}	&A\textunderscript{out}	&$\sigma$ 	&error &A\textunderscript{in}	&Z\textunderscript{in}	&A\textunderscript{out}	&$\sigma$ 	&error \\
 & & & [mb] & [mb] & & & & [mb] & [mb] \\
\hline
18	&6	&17	&10.2	&1.4	 &15	&5	&14	& 4.0	&1.0\\
18	&6	&15	&39.9	&3.2	 &15	&5	&12	&31.7	&2.5\\
18	&6	&14	&16.2	&1.8	 &15	&5	&11	&29.1	&2.7\\
18	&6	&13	&74.7	&5.3	 &15	&5	&10	&65.5	&5.5\\
18	&6	&12	&30.9	&3.0	 &15	&5	&9 	&10.8	&1.7\\
\hline
17	&6	&15	&27.9	&1.8	 &14	&5	&12	&21.3	&1.2\\
17	&6	&14	&14.1	&1.2	 &14	&5	&11	&20.6	&1.2\\
17	&6	&13	&72.5	&3.7	 &14	&5	&10	&62.8	&2.9\\
17	&6	&12	&40.9	&2.6	 &14	&5	&9 	&13.2	&1.0\\
17	&6	&11	&40.2	&2.5	\\
\hline
16	&6	&15	&20.5	&0.4	 &13	&5	&12	& 8.9	&0.3\\
16	&6	&14	&11.9	&0.3	 &13	&5	&11	&19.8	&0.5\\
16	&6	&13	&65.3	&1.0	 &13	&5	&10	&58.4	&1.1\\
16	&6	&12	&43.0	&0.7	 &13	&5	&9 	&17.6	&0.5\\
16	&6	&11	&53.7	&0.9	\\
16	&6	&10	& 4.1	&0.2	\\
\hline
15	&6	&14	&27.3	&1.2	 &12	&5	&11	& 6.8 	&0.3\\
15	&6	&13	&40.9	&1.6	 &12	&5	&10	&59.3	&1.6\\
15	&6	&12	&47.3	&1.8	 &12	&5	&9 	&20.6	&0.7\\
15	&6	&11	&67.7	&2.6	 &12	&5	&7 	& 3.5	&0.2\\
15	&6	&10	&10.4	&0.7	\\
\hline
14	&6	&13	&51.1	&1.4	 &11	&5	&10	&37.0	&1.3\\
14	&6	&12	&34.6	&1.1	 &11	&5	&9 	&19.9	&0.8\\
14	&6	&11	&84.8	&2.2	 &11	&5	&7 	& 3.0	&0.3\\
14	&6	&10	&16.7	&0.7	\\
\hline
13	&6	&12	&55.5	&1.3	 &10	&5	&9 	&13.3	&1.6\\
13	&6	&11	&76.2	&1.8	 &10	&5	&7 	&10.6	&1.6\\
13	&6	&10	&26.8	&0.9	\\
\hline
12	&6	&11	&85.4	&3.1	\\
12	&6	&10	&48.8	&2.2	\\
\hline
10	&6	&8 	&13.3	&3.0	\\

\hline
\hline
\end{tabular}
\caption{Summary of the extracted 1pxn removal cross sections. The error provided represents the statistical uncertainty. The systematic uncertainty due to uncertainties in the  target thickness and trigger efficiency is estimated to 5\%}
\label{table:crosssections}
\end{table}
Several isotopes were present in more than one fragment separator setting, and had therefore slightly different kinetic energies (390 $A$MeV to 430 $A$MeV). The cross sections at the slightly different energies did, as expected \cite{ref:ogawa2015}, not show any energy dependence in this interval and were averaged with respect to their statistical weights. The averaged cross sections are provided in Table~\ref{table:crosssections} and shown in Fig.~\ref{fig:xsecoverview}, which presents the production cross section versus $\Delta$A (difference in number of nucleons between mother and daughter nuclei) for incoming carbon and boron isotopes. 
For the latter we observe a strong trend in the production cross section of \textsuperscript{10}Be. It is the largest of all measured 1pxn cross sections for all isotopes for which the 1pxn removal leaves a Be-isotope with mass 10 or larger. For the carbon isotopes the trend is not as clear. Carbon isotopes lighter than mass 16 show clearly the largest 1pxn cross section for \textsuperscript{11}B, while those heavier than mass 15 have the largest cross section for semi-magic \textsuperscript{13}B. The transition point is \textsuperscript{16}C, featuring large production cross sections for both  \textsuperscript{11}B and \textsuperscript{13}B. A separate case is \textsuperscript{10}C which is proton-rich and for which only the 1p1n reaction populates a bound nucleus (\textsuperscript{8}B).

\section{Model calculations}
The model we use to understand the physics connected to our data is ABRABLA07 \cite{ref:gaimard1991}, which is a standard code for the description of fragmentation and fragmentation-fission reactions of heavy nuclei. It describes these reactions quite successfully (see e.g. Ref.~\cite{ref:helariutta2003}). Fragmentation is described by the model as a two-step process -- abrasion and ablation, the former determining how many nucleons are removed in the collision, and the latter which and how many light particles are evaporated  owing to the excitation energy induced by the collision. Both parts use the Monte-Carlo approach.\\
The abrasion part uses Karol's approximation \cite{Karol1975} to extract the total interaction cross section. The number of removed nucleons is calculated from the geometrical overlap of the colliding nuclei -- based on the impact parameter, while the neutron-proton ratio of the pre-fragment is calculated from the hyper-geometrical distribution \cite{ref:gaimard1991}. The excitation energy of the daughter nucleus is determined from the single-particle energies of the removed nucleons, which is on average 13.5 MeV per abraded nucleon \cite{ref:gaimard1991}. It was found \cite{ref:schmidt1993eefac} that the excitation energy has to be multiplied by a factor of two in order to reproduce experimental data, which is motivated by the final state interactions of participants and spectators.\\
The ablation part, described in detail in Ref.~\cite{ref:Kelic2009}, bases the particle emission on the statistical model and the Weisskopf-Ewing formalism \cite{ref:weisskopf1940}. Level densities are calculated using the Fermi-gas approach \cite{ref:gilbert1965fermigas}, modulated by nuclear structure effects (e.g. collective enhancement), which at low excitation energies is replaced by the constant-temperature model \cite{ref:Ignatyuk2001}.\\
Calculations were performed running 10\textsuperscript{6} collisions per incoming ion, rendering the statistical uncertainty of the calculated cross sections of 3 mb (the smallest experimental data point) to be below 2\%.

\section{Discussion}
\begin{figure}
\includegraphics[width=\columnwidth]{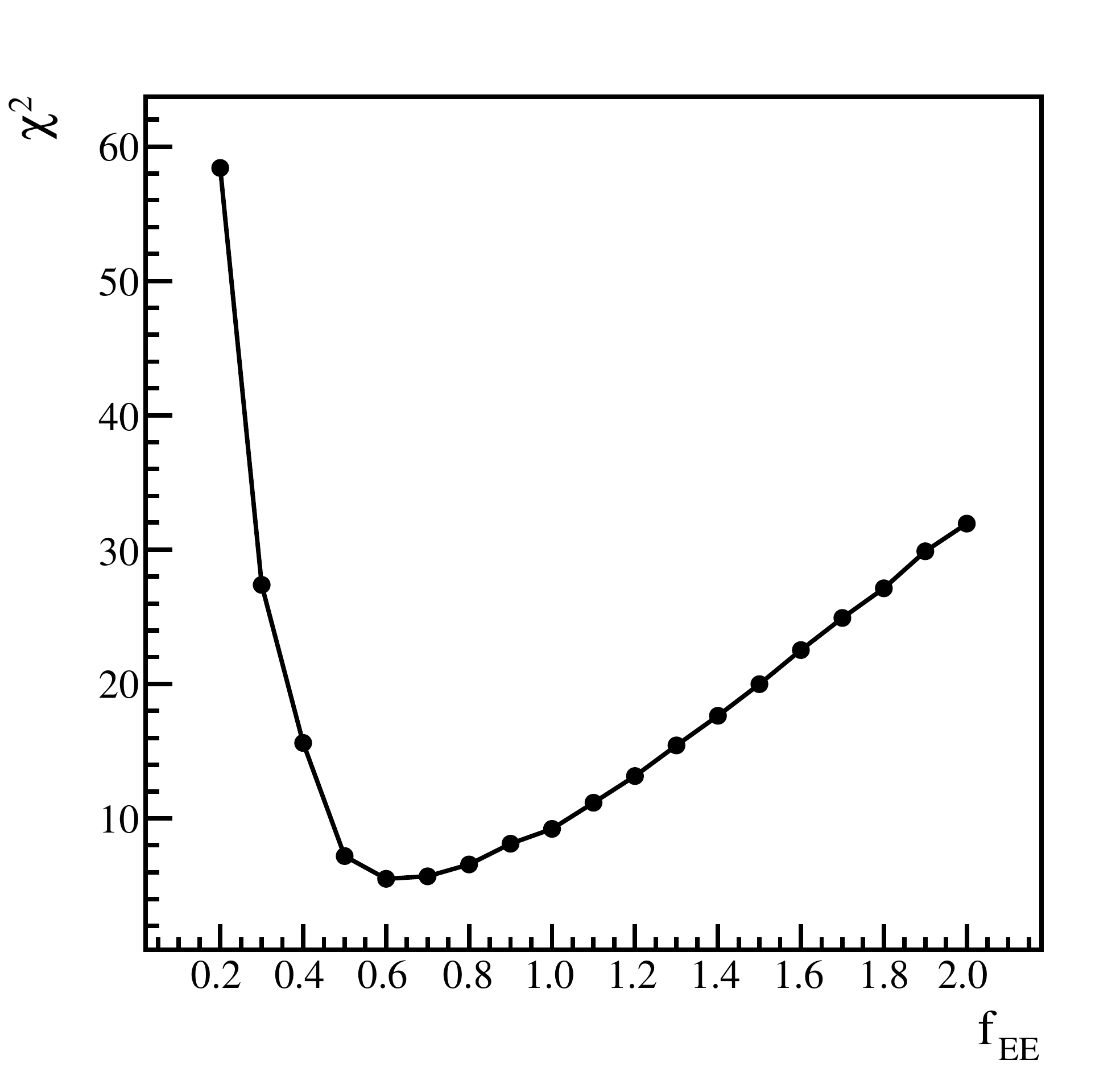}
\caption{Plot of the $\chi^2$ versus the excitation energy multiplication factor used in the ABRABLA07 \cite{ref:gaimard1991} calculations. The $\chi^2$ is determined as described in the text, summed for all experimentally determined cross sections measured in this work. Lines are used to guide the eye.}
\label{fig:eefacchi2}
\end{figure}

\begin{figure*}
\includegraphics[width=\textwidth]{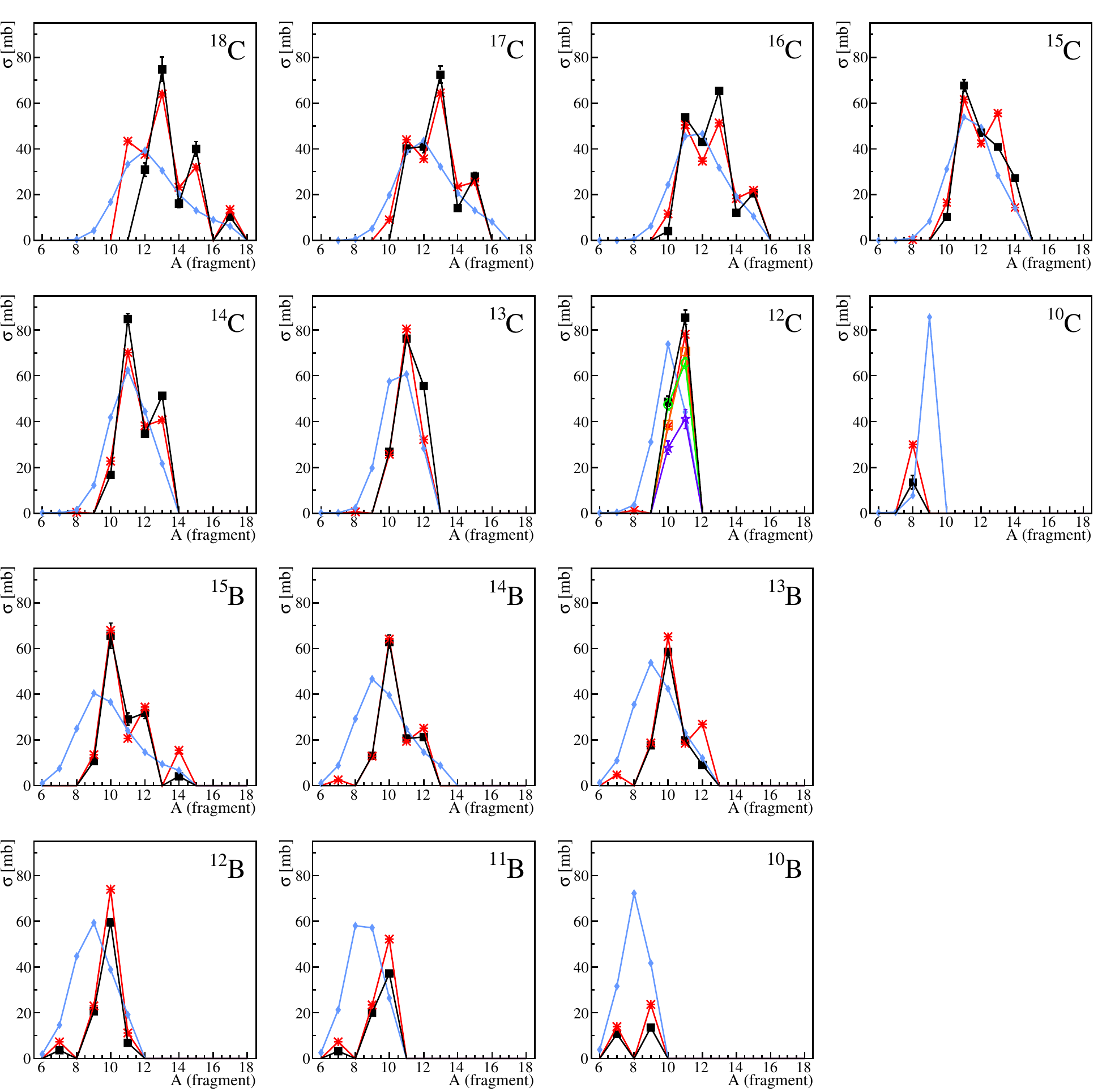}
\caption{(Colour online) Comparison between ABRABLA07 \cite{ref:gaimard1991} (red stars), EPAX \cite{ref:summerer2012, ref:summerer2013erratum} (blue diamonds) and the experimental data (black full squares). For \textsuperscript{12}C experimental data from three other measurements of \textsuperscript{12}C on C are shown: at 600 $A$MeV, Ref.~\cite{ref:webber1990} (orange empty square,); at 250 $A$MeV, Ref.~\cite{ref:kidd1988} (green empty circles); and at 400 $A$MeV, Ref.~\cite{ref:ogawa2015} (purple bold stars). }
\label{fig:xsecsurvey}
\end{figure*}

\begin{figure}
\includegraphics[width=\columnwidth]{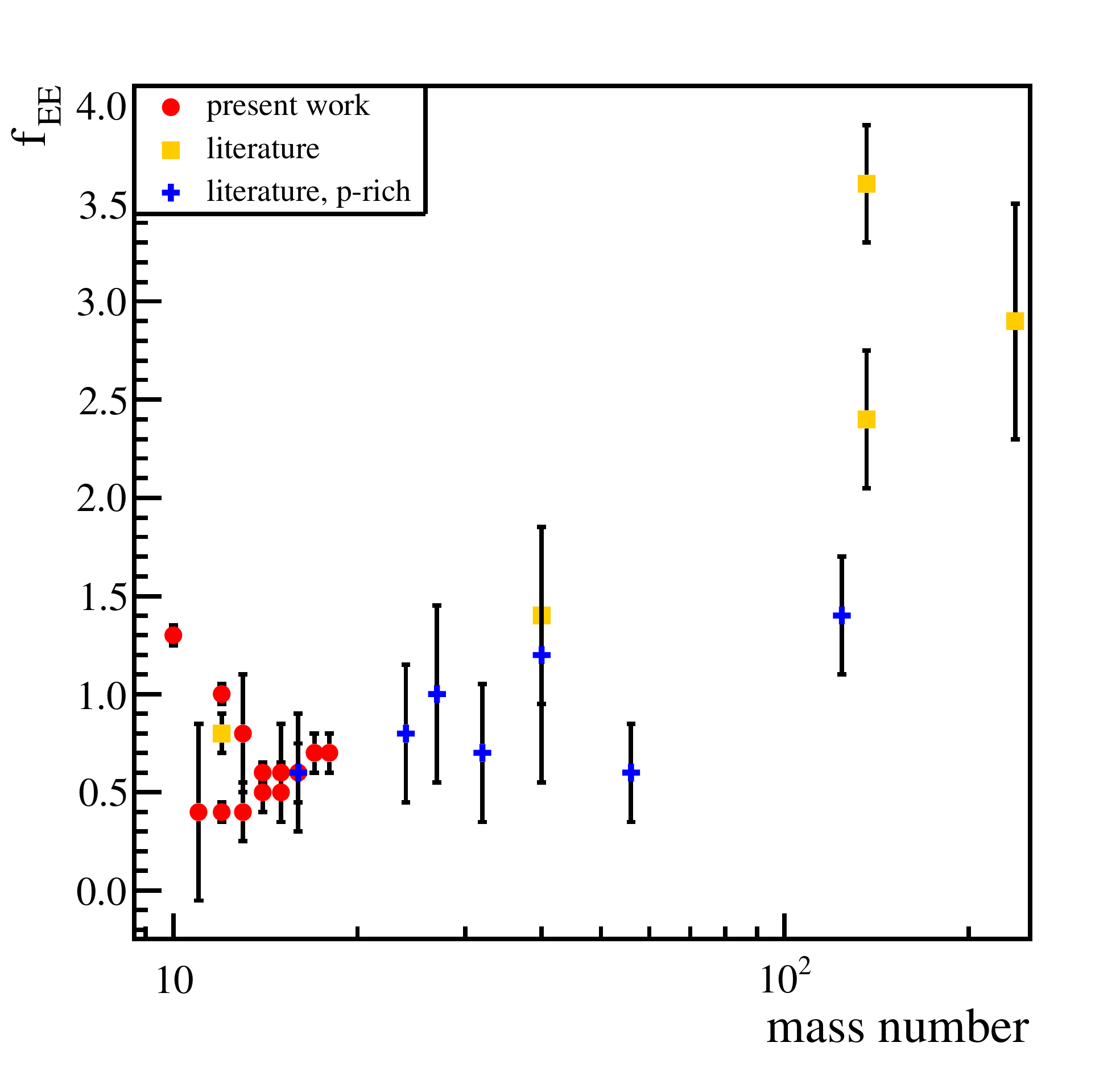}
\caption{ (Colour online) The optimal excitation energy multiplication factor versus the mass number. Error-bars indicate the estimated uncertainty, see text for details on the calculation. (Red) dots represent the present data, while (orange) squares indicate data from Ref.~\cite{ref:webber1990, ref:Nieto2007phd, ref:junghans1998, ref:henzlova2008, ref:Benlliure2008, ref:perez2011doublefrag}, and (blue) bold crosses represent data from Ref.~\cite{ ref:fernandez2005, ref:Fernandez2005erratum,ref:webber1990, ref:henzlova2008} for isotopes that have a larger neutron separation energy than proton separation energy. A clear difference between lighter and heavier nuclei is visible.}
\label{fig:eefac}
\end{figure}

To optimize the  input parameters of ABRABLA07,  we used the mass evaluation from 2012 \cite{ref:AME2012-1, ref:AME2012-2} instead of the mass evaluation from 2003 and added a few missing unbound nuclei. Both modifications resulted in very minor changes of the cross sections.\\
In order to be able to reproduce the cross sections of the light nuclei measured in this work, we had to decrease the multiplication factor of the excitation energy to 0.6.
This was deduced from a systematic study of the ability of ABRABLA07 to reproduce the  experimental cross sections depending on the excitation energy multiplication factor ($f_{EE}$). The study was performed by running ABRABLA07 calculations with an $f_{EE}$ varying between 0.2 and 2, in steps of 0.1. 
 Using both the statistic and known systematic uncertainty we calculated a $\chi^2$ for the agreement between calculation and data for each incoming isotope and $f_{EE}$.
 The result of the total $\chi^2$ per isotope, which is the sum of the individual $\chi^2$ of all incoming isotopes divided by the amount of daughter-isotopes, is illustrated in Fig.~\ref{fig:eefacchi2}. The minimum is located at 0.6, indicating that all isotopes simultaneously are best described by an $f_{EE}$ of 0.6, i.e. an average excitation energy of 8.1 MeV per abraded nucleon. \\
The complete comparison of the calculations with the best fit $f_{EE}$ (= 0.6) with the data is shown in Fig.~\ref{fig:xsecsurvey}. First, one should note that our experimental data for stable \textsuperscript{12}C agrees with data from previous stable beam experiments \cite{ref:webber1990,ref:kidd1988}. Data taken by Ogawa \textit{et al.} \cite{ref:ogawa2015} disagrees somewhat with both our and the other previous measurements.\\
Altogether, ABRABLA07, which is designed for calculation of fragmentation and fission cross sections of heavier nuclei, and employs several approximations based on the properties of these, reproduces the data very well. We still observe a few differences between model and data. Generally the prediction for 1pxn removal cross sections for B is much better than the prediction for 1pxn removal from C.
The 1p0n channels are generally overestimated for boron by ABRABLA07. For carbon no such trend is visible.\\
Another widely used model is EPAX developed by K. S\"ummerer \cite{ref:suemmerer2013epax}, which we also show for comparison (in Fig.~\ref{fig:xsecsurvey}). Our data is outside the range limit of EPAX being $A > 40$, but EPAX has previously been used for lower masses (e.g. Ref.~\cite{ref:leistenschneider2002}). This empirical formula misses details of the structure in this region of the nuclear chart, and has therefore only limited applicability for such light nuclei.\\
\begin{table}
\begin{tabular}{l@{\hskip 12pt} c}
\hline\hline
Reference & Isotope\\
\hline
this work & \textsuperscript{10}B, \textsuperscript{10}C, \textsuperscript{11}B, \textsuperscript{12}B, \textsuperscript{12}C, \textsuperscript{13}B, \textsuperscript{13}C,\\
&  \textsuperscript{14}B, \textsuperscript{14}C, \textsuperscript{15}B, \textsuperscript{15}C, \textsuperscript{16}C, \textsuperscript{17}C, \textsuperscript{18}C\\
\cite{ref:perez2011doublefrag} & \textsuperscript{132}Sn\textsuperscript{*}\\
\cite{ref:webber1990} & \textsuperscript{14}N\textsuperscript{*}, \textsuperscript{16}O, \textsuperscript{20}Ne\textsuperscript{*}, \textsuperscript{24}Mg, \textsuperscript{27}Al, \textsuperscript{28}Si\textsuperscript{*},\\
 &  \textsuperscript{32}S, \textsuperscript{40}Ar, \textsuperscript{40}Ca, \textsuperscript{56}Fe, \textsuperscript{58}Ni\textsuperscript{*}\\
\cite{ref:nieto2014} & \textsuperscript{208}Pb\textsuperscript{*}\\
\cite{ref:junghans1998} & \textsuperscript{238}U\\
\cite{ref:henzlova2008} & \textsuperscript{124}Xe, \textsuperscript{136}Xe\\
\cite{ref:Benlliure2008} & \textsuperscript{136}Xe\\

\cite{ref:fernandez2005,ref:Fernandez2005erratum} & \textsuperscript{92}Mo\textsuperscript{*}\\
\hline
\hline
\end{tabular}
\caption{Table summarizing which isotopes were used to study the mass dependence of the $f_{EE}$, sorted by publication. For isotopes marked with an asterix no minimal $\chi^2$ and therefore no optimal $f_{EE}$ could be determined.}
\label{table:literatureoverview}
\end{table}
A best fit $f_{EE} = 0.6$ for our data is quite different to the originally published $f_{EE}$ of 2.0 from peripheral collisions of the much heavier \textsuperscript{197}Au \cite{ref:schmidt1993eefac}. The final-state interactions, proposed as physics motivation for introducing the $f_{EE}$, should, from naive geometry arguments, scale with the size of the nuclei.
To further understand the influence of the excitation energy multiplication factor on the ability of ABRABLA07 to reproduce the 1pxn cross sections, we investigate the dependence of the $f_{EE}$ on the projectile mass. To do that we use data from Refs.~\cite{ref:webber1990, ref:Nieto2007phd, ref:junghans1998, ref:henzlova2008, ref:Benlliure2008, ref:perez2011doublefrag, ref:fernandez2005, ref:Fernandez2005erratum}, as summarized in Table~\ref{table:literatureoverview},
 and perform ABRABLA07 calculations with $f_{EE}$ between 0.5 and 4 in intervals of 0.1. With the requirements of beam energies above 100 $A$MeV and data available in tabulated form, we used all to our knowledge published 1pxn removal data available.\\
  For heavier isotopes, in contrast to light isotopes, the possibility of very long evaporation chains exists. These long evaporation chains are caused by reactions in which more excitation energy is generated in the abrasion step which corresponds to more violent, non-peripheral collisions. In order to compare similar collisions, we restrict ourselves to a maximum of 5 removed neutrons in this analysis, which corresponds to the same range as in our light nuclei.\\
  We calculate the $\chi^2$ (for each $f_{EE}$ and isotope), as above, which is then used to determine the best $f_{EE}$ for each isotope. For some isotopes no minimum could be found. This stems from a too large mismatch of the cross sections in our area of interest. The error is estimated by looking at which $f_{EE}$, other than the best, have a $\chi^2$ smaller than the best $\chi^2 + 1\sigma_{\chi^2}$. The error of the $\chi^2$ is estimated by standard error propagation. The largest possible difference between the $f_{EE}$ still having $\chi^2 \leq \chi^2_{best} + 1\sigma_{\chi^2,best}$ is determined for $f_{EE}$ being both smaller and larger than the best $f_{EE}$ and their average gives the estimated uncertainty.
  Large errors are caused by a mismatch between data and calculation concerning the trend of cross section vs. removed neutrons.\\
  Figure~\ref{fig:eefac} shows the best $f_{EE}$ versus mass number, for both our experimental data (red dots) and the data from literature (orange squares and blue bold crosses). Nuclei which have a smaller separation-energy for protons than for neutrons, which causes the particle-evaporation after the reaction to be different, are marked differently (blue bold crosses).  The figure shows that the excitation energy multiplication factor increases with increasing mass.\\
  Tarasov \textit{et al.} \cite{ref:tarasov2013} found, for fragmentation of \textsuperscript{82}Se at 139 $A$MeV, an excitation energy of 15 MeV per abraded nucleon with a different version of the abrasion-ablation model. This, though also central collisions are included, is consistent with our findings.  Unfortunately the region between mass 60 and mass 130 does not contain any data, so the transition from light to heavy masses is not very conclusive.\\
  Please note that the selection of the reaction channels (restriction to 1pxn with $0 \leq$ x $\leq 5$) included in our optimization of the $f_{EE}$, selects only peripheral reactions. This physics selection influences the result of the best fit $f_{EE}$, thus the results presented here are not in conflict with previous $f_{EE}$ = 2 results including the complete set of daughter nuclei.\\
    One can also observe that factors other than the mass  influence the induced average excitation energy, due to the large spread of the optimal $f_{EE}$ values. Concerning light nuclei, the description of the pre-fragment excitation energy in ABRABLA07 would benefit from improvement, since for these nuclei the influence of the nuclear structure and single-particle energies plays a bigger role. See e.g. Ref.~\cite{ref:Benlliure2015} for the importance of nuclear structure on pre-fragment excitation energy. Performing a simple test, decreasing the default potential depth in ABRABLA07 \cite{ref:gaimard1991} from 47.4 MeV to 40 MeV, we find no significant influence of that parameter on the ability of ABRABLA07 to reproduce our experimental data.\\

\section{Conclusions}
We have systematically measured 1pxn removal cross sections for 14 neutron-rich carbon and boron isotopes in one single experiment. These new data are used for comparison with model calculations. The  EPAX model deviates significantly from  the experimental data. The comparison of ABRABLA07 with the new data yields the necessity for a smaller average excitation energy in the model calculations for these nuclei. With that, the calculation reproduces the data surprisingly well, even though there are some deviations. Including additional data from literature we find that the average excitation energy in ABRABLA07 for best reproduction of experimental data on 1pxn ($0 \leq$ x $\leq 5$) reactions increases with increasing mass. This should be taken into account for future calculations of light nuclei with this model.\\
 However, the comparison to data also demonstrates that changing the average excitation energy per abraded nucleon alone is insufficient for a full description of the experimental data. The behaviour  of the induced excitation energy is complex, and more investigations are needed. A potential influence of the impact parameter on the $f_{EE}$, which is indicated by our results for heavy nuclei differing from the adopted value of $f_{EE}$ = 2, would be interesting to investigate further. A more realistic estimate of pre-fragment excitation energy would probably improve the model not only with regard to light isotopes, but more generally.\\
Due to its extreme relevance in helping us understand the isotope fragmentation production mechanism, we feel that additional theoretical improvements of the relatively successful abrasion-ablation model are necessary. In particular one deserves a better understanding and prediction of the average excitation energy per abraded nucleon.
\section*{Acknowledgements}
We would like to thank the FRS and the GSI accelerator staff for their efforts.
This work was supported by the Swedish Research Council. O. T. was supported by the Spanish research council under project FPA2012-32443, C. A. B. acknowledges support from the U.S. DOE grants DE-FG02-08ER41533 and the U.S. NSF Grant No. 1415656. L. M. F. acknowledges support by the Spanish research council under project FPA2013-41267-P and M. R. was supported by GSI (F\&E, DR-ZUBE), BMBF (06DR134I and 05P09CRFN5), the Nuclear Astrophysics Virtual Institute (NAVI, HGF VH-VI-417) and the Helmholtz Association Detector Technology and Systems Platform. T. K. acknowledges support by the German BMBF No. 06DA9040I, 05P12RDFN8 and 05P15RDFN1, while C. W. acknowledges funding by the UK STFC. 

\bibliographystyle{apsrev4-1} 
\bibliography{/home/ronja/Documents/mybib} 

\end{document}